\documentstyle[12pt]{article}

\def\p{{\rm {\bf p}}}
\def\q{{\rm {\bf q}}}
\def\Tr{{\rm Tr}}
\def\lg{{\rm log}}
\def\exp{{\rm exp}}
\def\g{\mbox{\boldmath$\gamma$}}

\begin{document}

\hfill BI-TP 96/55

\hfill November 1996

\vspace{1.5cm}

\begin{center}
{\bf FERMI EXCITATIONS IN HOT AND DENSE QUARK-GLUON PLASMA }
 \footnote{Research is partially supported by "Volkswagen-Stiftung"}

\vspace{1.5cm}
{\bf O.K.Kalashnikov}
\footnote{Permanent address: Department of
Theoretical Physics, P.N.Lebedev Physical Institute, Russian
Academy of Sciences, 117924 Moscow, Russia. E-mail address:
kalash@td.lpi.ac.ru}

Fakult\"at f\"ur Physik

Universit\"at Bielefeld

D-33501 Bielefeld, Germany

\vspace{2.5cm}

{\bf Abstract}
\end{center}

The Fermi excitations in hot and dense quark-gluon plasma are studied
in the Feynman gauge  using the temperature Green function technique.
We find the four well-separated branches for the case $m=0$  and
establish the additional splitting between them (the four different
masses) when $m\ne 0$. The long wavelength limit of these excitations
is found in the general case of the massive fermions at finite
temperature and densities to give the exact one-loop spectrum.
Simultaneously the many known results are reproduces as its
different limits.

\newpage

\section{Introduction}

At present the interest has essentially grown to study the quark
excitation spectra (in particular the quark masses) in hot QCD theory
at finite fermion densities. In the region of the large temperatures
and densities normally there are no massless particles (all quarks and
gluons, at least, have the dynamical masses $m\sim gT$ [1]) and these 
masses being not small influent qualitatively on many properties of the 
quark-gluon medium. Moreover the quarks being the Fermi particles are 
arranged under or near the Fermi sphere and the appearance of the 
diquark pairs is very probably  for a rather wide temperature region. 
On the other hand this coupling is not evident at once since the 
particle and hole excitations have the completely different effective 
masses and other properties that is especially pronounced for large 
$m$. There are a lot paper (starting from [2,3]) which in many aspects 
discuss the problem of the mass generations but today this scenario is 
well studied only in what concerns the dynamical quark mass which can 
be calculated perturbatively. The effective quark mass is usually 
calculated nonperturbatively to generate the chiral phase transition 
for hot and dense quark-gluon plasma and this question is open for 
discussion. Usually these investigations are rather complicated and 
frequently are accompanied by numerical calculations [4,5,6] where 
many of details are screening.

The goal of this paper is to find analytically the Fermi excitations
in hot and dense quark-gluon plasma and investigate their
peculiarities.  We use the standard temperature Green function
technique and fix the Feynman gauge for explicit calculations. The case
of a zero damping are only considered and we concentrate our attention
to solve the fermion dispersion relation in a more complete form. We
find the four well-separated branches for the case $m=0$  and establish
the additional splitting between them (the four different masses) when
$má\ne 0$. The long wavelength limit of these excitations is investigated
in the general case of the  massive fermions at finite temperature and
densities to give the exact one-loop spectrum.  We find the dynamical
quark masses (here only two values) and determine the four effective
quark masses which define the above mentioned splitting. This splitting
show that the particle excitations and holes have different masses and
that takes place for any $T,\mu$-values if bare $m\ne 0$.
Simultaneously a number of known results are reproduces as the
different limits of the solution found to demonstrate its reliability.

\section{ QCD Lagrangian and  quark self-energy}

The QCD Lagrangian in covariant gauges has the form
\setcounter{equation}{0}
\begin{eqnarray}
{\cal L}=&-&\frac{1}{4}{G_{\mu\nu}^a}^2+N_f{\bar \psi}
[\gamma_{\mu}(\partial_{\mu}-\frac{1}{2}
ig\lambda^aV_{\mu}^a)+m]\psi \nonumber\\ &-&\mu N_f{\bar
\psi}\gamma_4\psi +\frac{1}{2\alpha}(\partial_{\mu}V_{\mu}^a)^2 +{\bar
C}^a (\partial_{\mu}\delta^{ab}+gf^{abc}V_{\mu}^c)\partial_{\mu}C^b
\end{eqnarray}
where $G_{\mu\nu}^a=\partial_{\mu}V_{\nu}^a-\partial_{\nu}V_{\mu}^a
+gf^{abc}V_{\mu}^bV_{\nu}^c$  is the Yang-Mills field strength;
$V_{\mu}$ is a non-Abelian gauge field; $\psi$(and ${\bar
 \psi}$) are the quark fields in the SU(N)-fundamental representation
($\frac{1}{2}\lambda^a $ are its generators and $f^{abc}$ are the
SU(N)-structure constants) and $C^a$ (and ${\bar C}^a$) are the ghost
Fermi fields. In Eq.(1) $\mu$ and $m$ are the quark chemical potential
and bare quark mass, respectively, $N_f$ is the number of quarks
flavours and $\alpha$ is the gauge fixing parameter ($\alpha=1$ for
the Feynman gauge). The metric is chosen to be Euclidean and
$\gamma_{\mu}^2=1$.

Our starting point is the exact Schwinger-Dyson equation for the
temperature quark Green function
\begin{eqnarray}
G^{-1}(q)=G_0^{-1}+\Sigma(q)
\end{eqnarray}
where the quark self-energy has the standard representation
\begin{eqnarray}
\Sigma(q)=\frac{N^2-1}{2N}\frac{g^2}{\beta}\sum_{p_4}\int\frac{d^3p}
{(2\pi)^3}{\cal D}_{\mu\nu}(p-q)\gamma_{\mu}G(p)\Gamma_{\nu}(p,q|p-q)
\end{eqnarray}
Here we calculate $\Sigma$ only in the one-loop approximation
using the bare Green functions in Eq.(3) and fixing the Feynman gauge.
The bare functions in accordance with Eq.(1) have the form
\begin{eqnarray}
\Gamma_{\mu}^0(p,q|p-q)=\gamma_{\mu}\,,\qquad G^0(p)=
\frac{-i\gamma_{\mu}{\hat p_{\mu}}+m}{{\hat p^2}+m^2}
\,,\qquad {\cal D}_{\mu\nu}^0(p)=\frac{\delta_{\mu\nu}}{p^2}
\end{eqnarray}
where ${\hat p}=\{(p_4+i\mu),\p \}$ is the convenient notation.
The summation over the spinor indices are performed within
Eq.(3) using the $\gamma$-matrix algebra and the result is found to be
\begin{eqnarray}
\Sigma(q)=\frac{N^2-1}{N}\frac{g^2}{\beta}\sum_{p_4}^F\int\frac{d^3p}
{(2\pi)^3}\;\frac{i\gamma_{\mu}{\hat p_{\mu}}+2m}{({\hat
p^2}+m^2)\;(p-q)^2}
\end{eqnarray}
Now we introduce two new functions  and rewrite Eq.(5) as follows
\begin {eqnarray}
\Sigma(q)=i\gamma_{\mu}K_{\mu}(q)+m\;Z(q)
\end{eqnarray}
where $K_\mu(q)=q_\mu\;a(q)+iu_\mu\;b(q)$. Here $u_\mu=\{1,0\}$ is the
standard medium vector and all functions separately depend on $q_4$
an $|\q|$. The decomposition (6) is not completed [3,7] but all other
functions are not generated in the one-loop approximation. Using
decomposition (6) we transform Eq.(2) into the form
\begin{eqnarray}
G(q)=\frac{-i\gamma_{\mu}({\hat q_{\mu}}+K_{\mu})+m\;(1+Z)}
{({\hat q_{\mu}}+K_\mu)^2\;+\;m^2\;(1+Z)^2}
\end{eqnarray}
which is more convenient futher. Setting the determinant of Eq.(7) to
zero, we find the dispersion equation
\begin{eqnarray}
({\hat q_{\mu}}+K_{\mu})^2\;+\,m^2\;(1+Z)^2=0
\end{eqnarray}
which determines the Fermi excitation spectra after the standard
analytic continuation.

\section{ Calculations of the quark self-energy}
Now the summation over the Fermi frequencies $p_4=2\pi T(n+1/2)$
in Eq.(5) is performed with the aid of the usual prescription [8] and
all terms are collected in the convenient form using the simple
algebraic trasformations.  The result is given by
\begin{eqnarray}
&&\!\!\!\!\!\!\!\!\!\!\!\!\!\!\!\Sigma(q)=-\frac{g^2(N^2-1)}{N}
\int\frac{d^3p}{2(2\pi)^3}\;\left\{\;\Bigr[
\frac{1}{\epsilon_\p}\;\frac{n_\p^+\;[\gamma_4\epsilon_\p+(i
\g\p+2m)]}
{[q_4+i(\mu+\epsilon_\p)\;]^2+(\q-\p)^2}\right.\nonumber\\
&&\!\!\!\!\!\!\!\!\!\!\!\!\!\!\!\!\!\! \left.+\;\frac{n_\p^B}{|\p|}
\;\frac{(|\p|+\mu-iq_4) \gamma_4-[i
\g(\q-\p)+2m]}{[q_4+i(\mu+|\p|)\;]^2+\epsilon_{\p-\q}^2}\;\Bigr]
-\Big[h.c.(m,\mu)\rightarrow-(m,\mu)\Big]\right\}
\end{eqnarray}
where $\epsilon_\p=\sqrt{\p^2+m^2}$ is the bare quark energy;
$n_\p^{B}=\left\{\exp\beta|\p|-1\right\}^{-1}$ and $n_\p^{\pm}=
\left\{\exp\beta(\;\epsilon_\p \pm \mu)+1\right\}^{-1}$ are the Bose
and Fermi occupation number, respectively.

Before the last integration over the angles being performed it is
desirable to separate the functions $K(q)$ and $Z(q)$ to simplify all
further calculations. Using that $\Tr \Sigma(q)/4=m\;Z(q)$ we
find the function $Z(q)$
\begin{eqnarray}
&&Z(q)=-\frac{g^2(N^2-1)}{N}\int\frac{d^3p}
{(2\pi)^3}\;\left\{\;\Bigr[\;\frac{1}{\epsilon_\p}\;\frac{n_\p^+}
{[q_4+i(\mu+\epsilon_\p)\;]^2+(\q-\p)^2}\right.\nonumber\\
&&-\left.\frac{n_\p^B}{|\p|}\;\;\frac{1}
{[q_4+i(\mu+|\p|)\;]^2+\epsilon_{\p-\q}^2}\;
\Bigr]\;+\;\Big[\;h.c.(\mu\rightarrow-\mu)\;\Bigr]\;\right\}
\end{eqnarray}
and then analogously we find the vector $K_\mu(q)$. At first,
$\Tr \gamma_4 \Sigma(q)/4=iK_4(q)$  reproduces the
$K_4$-function
\begin {eqnarray}
&&iK_4(q)=-\frac{g^2(N^2-1)}{N}\int\frac{d^3p}
{2(2\pi)^3}\;\left\{\;\Bigr[\;\frac{n_\p^+\;}{[q_4+i(\mu
+\epsilon_\p)\;]^2+(\q-\p)^2}\right.\nonumber\\
&&\left.+\frac{n_\p^B}{|\p|}\;\frac{|\p|+\mu-iq_4}
{[q_4+i(\mu+|\p|)\;]^2+\epsilon_{\q-\p}^2}\;\Bigr]\;
-\;\Bigr[\;h.c.(\mu\rightarrow-\mu)\;\Bigr]\;\right\}
\end{eqnarray}
and then $\Tr \gamma_n \Sigma(q)/4=iK_n(q)$ gives the
vector $K_n$ (n=1,2,3)
\begin {eqnarray}
&&K_n(q)=-\frac{N^2-1}{N}g^2\int\frac{d^3p}{2(2\pi)^3}\;\left\{\;\Bigr[
\;\frac{1}{\epsilon_\p}\;\frac{n_\p^+\;p_n}{[q_4+i(\mu+\epsilon_\p)\;]^2
+(\q-\p)^2}\right.\nonumber\\
&&\left.-\frac{n_\p^B}{|\p|}\;\;\frac{(\q-\p)_n}{[q_4+i(\mu
+|\p|)\;]^2+\epsilon_{\q-\p}^2}\;\Bigr]\;+\;\Big[\;h.c.(\mu\rightarrow
-\mu)\;\Big]\;\right\}
\end{eqnarray}

All the further calculations are simple but rather complicated when
the integration over angles within Eqs.(10)-(12) is performed.
Here we use the two standard integrals and introduce the new notations
to simplify the final result. These notations are as follows
\begin{eqnarray}
a_F^{\pm}&=&\frac{\q^2-m^2-(iq_4-\mu)^2 \pm
2\epsilon_\p(iq_4-\mu)-2|\p||\q|} {\q^2-m^2-(iq_4-\mu)^2 \pm
2\epsilon_\p(iq_4-\mu)+2|\p||\q|}
\end{eqnarray}
for integrals with the fermion distribution function  and
analogously
\begin{eqnarray}
a_B^{\pm}&=&\frac{\q^2+m^2-(iq_4-\mu)^2
\pm 2|\p|(iq_4-\mu)-2|\p||\q|} {\q^2+m^2-(iq_4-\mu)^2 \pm
2|\p|(iq_4-\mu)+2|\p||\q|}
\end{eqnarray}
and for integrals with boson one.

After all algebra is made the function $Z(q)$ has the form
\begin{eqnarray}
&&Z(q_4,\q)=\frac{g^2(N^2-1)}{N}\int\limits_{0}^{\infty}
\frac{d|\p|}{4\pi^2}\;\frac{|\p|}{2|\q|}\left\{\;\frac{1}{\epsilon_\p}
\Bigr[\frac{n_\p^++n_\p^-}{2}\;\lg(a_F^+a_F^-)\right.\nonumber\\
&&\left.+\frac{n_\p^+-n_\p^-}{2}\;\lg(\frac{a_F^+}{a_F^-})
\Bigr]-\frac{n_\p^B}{|\p|}\;\lg(a_B^+a_B^-)\right\}
\end{eqnarray}
and the similar form has the function $K_4(q)$
\begin {eqnarray}
&&\!\!\!\!\!\!\!\!\!\!iK_4(q_4,\q)=\frac{g^2(N^2-1)}{N}\int\limits_
{0}^{\infty}\frac{d|\p|}{8\pi^2}\frac{|\p|}{2|\q|}\left\{\;\Bigr[
\;\frac{n_\p^++n_\p^-}{2}\;\lg(\frac{a_F^+}{a_F^-})\right.\nonumber\\
&&\!\!\!\!\!\!\!\!\!\!\left.+\frac{n_\p^+-n_\p^-}{2}\;
\lg(a_F^+a_F^-)\;\Bigr]+n_\p^B\;\Bigr[\;\lg(\frac{a_B^+}{a_B^-})+
\frac{(\mu-iq_4)}{|\p|}\;\lg(a_B^+a_B^-)\Bigr]\right\}
\end{eqnarray}
The more complicated calculations are necessary to obtained the
vector $K_n(q)$ where $n=1,2,3$. Here we use a definition
$K_n(q)=q_n K(q)$ and after that the scalar K-function will be
calculated. The result has the form
\begin {eqnarray}
&&\!\!\!\!\!\!\!q^2\;K(q_4,\q)=\frac{g^2(N^2-1)}{N}
\int\limits_{0}^{\infty}\frac{d|\p|}{4\pi^2}\left\{\;\frac{\p^2}
{\epsilon_\p}\Biggr(\;\frac{n_\p^+-n_\p^-}{2}\;\Bigr[\;\frac{1}
{8|\p||\q|}\;\Bigr(h_F\ln(\frac{a_F^+}{a_F^-})\right.\nonumber\\
&&\!\!\!\!\!\!\!\!\left.+d_F\ln(a_F^+a_F^-)\Bigr)\Bigr]+\frac{n_\p^+
+n_\p^-}{2}\;\Bigr[\;1+\;\frac{1}{8|\p||\q|}\Bigr(\;h_F\ln(a_F^+a_F^-)
+d_F\ln(\frac{a_F^+}{a_F^-})\;\Bigr)\Bigr]\;\Biggr)\right.\nonumber\\
&&\!\!\!\!\!\!\!\!\left.+n_\p^B\;|\p|\;\bigr[1+\frac{1}{8|\p||\q|}
\Bigr(h_B\;\ln(a_B^+a_B^-)+d_B\;\ln(\frac{a_B^+}
{a_B^-})\;\Bigr)-\frac{|\q|}{4|\p|}\ln(a_B^+a_B^-)\Bigr]\right\}
\end{eqnarray}
where $h_F=\q^2-m^2-(iq_4-\mu)^2$ and $d_F=2\epsilon_\p(iq_4-\mu)$ ( the
analogously $h_B=\q^2+m^2-(iq_4-\mu)^2$ and $d_B=2|\p|(iq_4-\mu)$)
are the new notations.

\section{ The quark excitation spectrum}
Here we use Eq.(8) to find the spectrum of the Fermi excitations of hot
quark-gluon plasma in the presence of background for the different
values of the $T$, $\mu$ and $m$ parameters. The cases of a zero
damping are only considered and due to this fact our analytical
continuation is trivial.

\subsection{The spectrum for all $|\q|$ in the case $m=0$}
This is a more simple case which can be exactly considered in the
analytic form. Here we find the Fermi excitation spectrum for all $|q|
\ne 0$ within the standard high temperature technique keeping $ \mu \ne
0$.  To this end we expand Eqs.(15)-(17) taking only the leading
$T^2$-term with the $\mu/T$-corrections and solve the dispersion
equation selfconsistently exploiting everywhere the new mass shell.
Our starting point is the expressions for $a_F^{\pm}$ and $a_B^{\pm}$
quantities which being equal for the case $m=0$ determine all functions
which are necessary to expand. Using the integration of the $a^{\pm}$
quantities over $|p|$ we replace $|\p|$ to $|\p|T$ and find the
following expansions
\begin {eqnarray}
\ln(a^+a^-)=-\frac{2|\q|}{|\p|}+O(\frac{1}{T^2})\,,\qquad
\ln\frac{a^+}{a^-}=2\ln\frac{\xi-1}{\xi+1} + O(\frac{1}{T^2})
\end{eqnarray}
which are able to simplify essentially the further calculations. Here
$\xi=(iq_4-\mu)/|\q|$ is a convenient variable.

For our case the dispersion equation (8) can be presented as follows
\begin {eqnarray}
[\;(iq_4-\mu)-{\bar K}_4]^2\;=\;\q^2\;(1+K)^2
\end{eqnarray}
and at once to solve it as usual
\begin {eqnarray}
(iq_4-\mu)-{\bar K}_4\;=\;\eta |\q|\;(1+K)
\end{eqnarray}
where $\eta=\pm 1$ and we use the definition $K_4=i{\bar K}_4$.  All
functions within Eq.(20) should be calculated with the aid of the
expansions (18) before we solved Eq.(20) explicitly.

These calculations are easily performed and one finds the following
result
\begin {eqnarray}
K(q_4,\q)\;=\;\frac{I_K}{\q^2}\Bigr(\;1+\frac{\xi}{2}\ln
\frac{\xi-1}{\xi+1}\Bigr)\;+\;I_B\;\Bigr(\;\xi-\frac{1}{2}
(1-\xi^2)\ln\frac{\xi-1}{\xi+1}\;\Bigr)
\end{eqnarray}
for the function $K$ and analogously for another function
\begin {eqnarray}
-{\bar K}_4(q_4,\q)\;=\;\frac{I_K}{2|\q|}
\ln\frac{\xi-1}{\xi+1} +I_B
\end{eqnarray}
Here the integrals are defined to be
\begin {eqnarray}
&&I_K\;=\;\frac{g^2(N^2-1)}{N}\int\limits_0^{\infty}\frac{d|\p|}{4\pi^2}
\;|\p|\;\Bigr[\;\frac{n_\p^++n_\p^-}{2}\;+\;n_\p^B\;\Bigr] \\
&&I_B=-\frac{g^2(N^2-1)}{N}\int\limits_0^{\infty}
\frac{d|\p|}{4\pi^2}\;\frac{n_\p^+-n_\p^-}{2}
\end{eqnarray}
Now one should put the expressions found above into Eq.(20) and
perform a number of the algebraic transformations to find
$\omega=\xi|\q|$. Here $\omega=(iq_4-\mu)$ and our variable is $\xi$
which is a more convenient then  $|\q|$. The latter should be excluded
with the aid of Eq.(20).  The result is the quadratic equation with
respect to $\omega$
\begin {eqnarray}
\omega^2\Bigr(\xi-\eta[1+I_B\;B(\xi)]\;\Bigr)+\omega\;\xi
I_B+I_K\;\xi^2A(\xi)=0
\end{eqnarray}
where the functions $A(\xi)$ amd $B(\xi)$ are given by
\begin {eqnarray}
A(\xi)=\eta \;(1+\frac{\xi-\eta}{2}\ln\frac{\xi-1}{\xi+1}\;)\nonumber\\
B(\xi)=\xi-\frac{1}{2}(1-\xi^2)\ln\frac{\xi-1}{\xi+1}
\end{eqnarray}
Keeping the accuracy of calculations our solution of Eq.(25) is found
to be
\begin {eqnarray}
E(\xi)=\mu-\frac{\xi\;I_B}{2(\xi-\eta)}\;
\pm\xi\sqrt{I_K\;\Bigr(\;\frac{\eta}
{\xi-\eta}+\frac{\eta}{2}\ln\frac{\xi-1}{\xi+1}\;\Bigr)}
\end{eqnarray}
which presents the four branches of the Fermi excitations in the medium.
Here $\eta=\pm 1$ and we return the physical variable $E=ip_4$. Eq.(27)
is a more general one-loop result for the case $m=0$ and contains all
known ones. For example, if $\mu=0$, the well-known spectrum found in
paper [2]
\begin {eqnarray}
\omega_\pm^2(\xi)=\xi^2\;\omega_0^2\;\Bigr(\;
\frac{\eta} {\xi-\eta}+\frac{\eta}{2}\ln\frac{\xi-1}{\xi+1}\;\Bigr)
\end{eqnarray}
is in agreement with Eq.(27) if one puts $\mu=0$ and squares it. In
Eq.(28) $\omega_0^2=g^2T^2/6$ as it is given by Eq.(23) for $N=3$.
Within Eq.(27) $1<\xi<\infty$ and the long wavelength limit corresponds
to $\xi\rightarrow\infty$. For this limit one finds the very simple
formula
\begin{eqnarray} E=\mu-\frac{I_B}{2}\;\pm\;\sqrt{I_K}
\end{eqnarray}
which, indeed, is correct for any $T$ (and for the case $T=0$ as well).
We demonstrate this interesting fact in the next section although, in
principle, this situation is known.  Eq.(29) represents the more
general expression for the dynamical quark mass when $m=0$
\begin {eqnarray}
M^2\;=\;\frac{g^2(N^2-1)}{N}\int\limits_0^{\infty}\frac{d|\p|}{4\pi^2}
\;|\p|\;\Bigr[\;\frac{n_\p^++n_\p^-}{2}\;+\;n_\p^B\;\Bigr]
\end{eqnarray}
which for many special cases is well-known. For example,
when $T=0$  one finds the dynamical mass as follows [9]
\begin {eqnarray}
M^2\;=\;\frac{N^2-1}{N}\frac{g^2\mu^2}{16\pi^2}
\end{eqnarray}
and in another limit $\mu=0$ this expression was calculated in paper
[2].

\subsection{The long wavelength limit with $m \ne 0$}
This is a rather important limit which determines the effective quark
mass taking into account all radiative corrections: here only the
one-loop corrections. We present our result keeping the perturbative
accuracy and establish the additional splitting between the branches
found above.

Our starting point is Eq.(8) which for the case $m \ne 0$ it is
useful to rewrite as follows
\begin{eqnarray}
iq_4(1+a)\;=\;\mu+b +\eta \sqrt{\;m^2(1+Z)^2\;+\;\q^2(1+a)}
\end{eqnarray}
where $\eta=\pm 1$ and all functions should be calculated by using
Eq.(6) and its decomposition.  These calculations yield the simple
result \begin{eqnarray} a(q)=\frac{(qK(q))\;-\;(uq)(uK(q))}{q^2-(uq)^2}
\,,\qquad ib(q)=(uK(q))-(qu)a(q)
\end{eqnarray}
and now Eq.(32) can be transformed in the form
\begin{eqnarray}
i\q_4\;=\;\mu\;-i(uK(q))\;+\eta\;\sqrt{m^2(1+Z)^2+\q^2+(\q{\bf
K}(q))}
\end{eqnarray}
which is convenient for the futher calculations
in the case studied.  Below Eq.(34) will be presented in a more
explicit form in the long wavelength limit.

If the limit $|\q|=0$ is only considered all functions which determine
Eq.(34) can be simplify by doing the necessary expansions within
Eqs.(15)-(17) or exploiting directly Eqs.(10)-(12) in the
$|\q|=0$ limit. After all algebraic transformations being made all
terms are collected to reproduce the final result in the convenient
form.

The function $K_4(q_4,0)$ is found to be
\begin {eqnarray}
&&\!\!\!\!\!\!-i(uK)(q_4,0)\;=\;\frac{g^2(N^2-1)}{N}\int
\limits_0^{\infty}\frac{d|\p|}{4\pi^2}\left\{\frac{4\p^2\epsilon_\p}
{4\epsilon_\p^2-(iq_4-\mu)^2[1+{\displaystyle\frac{m^2}
{(iq_4-\mu)^2}}]^2}\right.\nonumber\\
&&\!\!\!\!\!\!\left.\Bigr[\;\frac{n_\p^++n_\p^-}{2}\;
+\;\frac{(iq_4-\mu)}{2\epsilon_\p}\Bigr(1+\frac{m^2}
{(iq_4-\mu)^2}\Bigr)\frac{n_\p^+-n_\p^-}{2}\Bigr]\right.\\
&&\!\!\!\!\!\!\left.+n_\p^B\;\frac{4\p^3}{4\p^2-(iq_4-\mu)^2
[1-{\displaystyle\frac{m^2}{(iq_4-\mu)^2}}]^2}\Bigr[\;1+\frac{(iq_4
-\mu)^2-m^2}{2|\p|^2}\Bigr]\right\}\;\frac{1}{iq_4-\mu}\nonumber
\end{eqnarray}
and the function $Z(q_4,0)$ which determines the mass renormalization
can be presented a follows
\begin {eqnarray}
&&Z(q_4,0)\;=\;-\frac{g^2(N^2-1)}{N}\int\limits_0^{\infty}
\frac{d|\p|}{4\pi^2}\;\left\{\;\frac{4\p^2}
{4\epsilon_\p^2(iq_4-\mu)^2-[(iq_4-\mu)^2+m^2]^2}\right.\nonumber\\
&&\left.\Bigr[\;\frac{(iq_4-\mu)^2+m^2}
{\epsilon_\p}\;\frac{n_\p^++n_\p^-}{2}
\;+\;2(iq_4-\mu)\;\frac{n_\p^+-n_\p^-}{2}\Bigr]\right.\nonumber\\
&&\left.-\;n_\p^B\;\frac{4|\p|\;[(iq_4-\mu)^2-m^2]}
{4|\p|^2(iq_4-\mu)^2-[(iq_4-\mu)^2-m^2]^2}\right\}
\end{eqnarray}
The more complicated calculations yield that $(\q{\bf K})(q_4,0)=0$

Now our problem is to solve Eq.(34) explicitly. To this end we should
put Eq.(35)-(36) on the new mass shell $(iq_4-\mu)=\omega$ and in this
form substitute these integrals into Eq.(34). However in this case the
arisen equation is very complicated and can be used only for numerical
calculations [4]. On the other hand keeping the perturbative accuracy
the obtained integrals are possible to simplify considering the quantity
$\omega^2-m^2\approx 0$ inside them.  In doing so, the result has the
form
\begin{eqnarray}
-iK_4(q_4,0)\;=\;\frac{I_A}{\omega}\;-\;I_B\\
Z(q_4,0)\;=\;-2I_Z\;+\;2\frac{I_B}{\omega}
\end{eqnarray}
where the new integrals are:
\begin {eqnarray}
&&I_A\;=\;\frac{g^2(N^2-1)}{N}\int\limits_0^{\infty}\frac{d|\p|}{4\pi^2}
\;\Bigr[\;\epsilon_\p\;\frac{n_\p^++n_\p^-}{2}\:+\;|\p|\;n_\p^B
\;\Bigr]\nonumber\\
&&I_Z=\frac{g^2(N^2-1)}{N}\int\limits_0^{\infty}
\frac{d|\p|}{4\pi^2}\;\frac{n_\p^++n_\p^-}{2\epsilon_\p}
\end{eqnarray}
and $I_B$ was earlier determined by Eq.(24). Now the dispersion
equation is the simple quadratic equation
\begin{eqnarray}
\omega^2-\omega [\;\eta\; m (1-2I_Z)-I_B\;]-(\;I_A+2\eta mI_B\;)\;=\;0
\end{eqnarray}
whose solution reproduces the final result. This result has the form
\begin {eqnarray}
\omega=\frac{1}{2}\Big[\eta\;m(1-2I_Z)-I_B\Big]\pm
\sqrt{\;\frac{[\eta\;m(1-2I_Z)-I_B]^2}{4} +(I_A+2\eta mI_B)}
\end{eqnarray}
which gives the additional splitting of the branches found above. This
splitting show that the particle excitations and holes have different
masses and that takes place for any $T,\mu$-values if the bare mass
$m\ne 0$.  In Eq.(41) all parameters are free and this solution
extends the previously known result [6] taking into account all
one-loop corrections. The dynamical quark mass is defined to be
\begin {eqnarray}
M^2\;=\;I_A+2\eta mI_B
\end{eqnarray}
and this expression is the more general one-loop result for this
quantity. The quark mass shift has a rather complicated form
\begin {eqnarray} \delta
m=-\frac{1}{2}\Big[\eta m(1+2I_Z)+I_B\Big]\pm \sqrt{\frac{[\eta
m(1-2I_Z)-I_B]^2}{4} +(I_A+2\eta mI_B)}
\end{eqnarray}
and is different for each branches as it mentioned above. This means
that a question about phenomenological quark mass remains open and
requires the additional studying.

In the limit $T=0$ all integrals in Eq.(41) are exactly calculated.
For example within the standard QCD (where $N=3$) one has
\begin{eqnarray}
&&I_Z=\frac{g^2}{3\pi^2}\ln\frac{\mu+\sqrt{\mu^2-m^2}}{m}\;;\;\;
I_B=\frac{2g^2}{3\pi^2}\sqrt{\mu^2-m^2}\nonumber\\ { }\nonumber\\
&&I_A=\frac{g^2}{6\pi^2}\Bigr(\;\mu\sqrt{\mu^2-m^2}
\;+\;m^2\ln\frac{\mu+\sqrt{\mu^2-m^2}}{m}\;\Bigr)
\end{eqnarray}
and the spectrum  can be investigated by using the simple algebraic
transformations.  In particular case when $m=0$ the spectrum has a
rather simple form
\begin {eqnarray} E=\mu(1-\frac{g^2}{3\pi^2})\pm
\sqrt{\;\frac{g^2\mu^2}{6\pi^2}}
\end{eqnarray}
which presents the one-loop result for the massless quark excitation
in the cold medium. In the case $m=0$  Eq.(41) coincides with Eq.(29)
as it was promised.  The found equality proves Eq.(29) and this remark
is very useful for the practical calculations.

\section{Conclusion}

To summarize we have solved the fermion dispersion relation in the
general case of the massive fermions  at finite temperature and
densities. We find that Fermi excitations in medium for $m=0$ have
the four well-separated branches: two of them present the particle
excitations and two other correspond to the antiparticles ones. The
additional splitting of the branches mentioned above is found for the
case $m\ne 0$  which demonstrate  that the effective mass for all
branches are different for any $T,\mu$-values if only bare $m\ne 0$.
The dynamical mass have only two different values in the general case
$m\ne 0$ and an unique value in other cases. Here we see the problem
with the definition of the phenomenological quark mass since there is a
question which branch it presents. It is also essential that the
particles excitations and the hole ones have a different behaviour with
respect of the bare mass and there is a problem to find any
correlations between them . Another problem concerns the gauge
dependence of all results found here and in other papers (see e.g
[10]). This is a serious problem and it is not excluded that a number
of the one-loop results are, indeed, gauge dependent. In the first
place it concerns a damping (where resummation is necessary [11])
and probably all the next-to-leading order terms where the
same scale $\sim g^2T$ is essential. However, in our opinion the
long wavelength limit of any excitation spectra when the damping
is not important should be gauge independent. This statement is 
in agreement with [12] but there is a different result [13] where even 
$T^2$-term is gauge dependent. So today the problem remains and in 
this situation the Feynman gauge used here is a more reliable one to 
have correct result.  This problem is still pronounced for many 
nonperturbative calculations which usually use the axial gauge (see 
e.g.[5,14]). Moreover at present the peculiarities of many singular 
gauges are exploited to produce the physical results [15]. However 
these calculations being very complicated requires the special 
attention and should be discussed in a separate paper.

\begin{center}
{\bf Acknowledgments}
\end {center}
I would like to thank Rudolf Baier for comments of this paper and
all the colleagues from the Department of Theoretical Physics of the
Bielefeld University for the kind hospitality.

\newpage

\begin{center}
{\bf References}
\end{center}

\renewcommand{\labelenumi}{\arabic{enumi}.)}
\begin{enumerate}

\item{ O.~K.~Kalashnikov and V.~V.~Klimov, Sov. J. Nucl. Phys.  {\bf
31} (1980) 699 ( Yad. Fiz. {\bf 31} (1980) 1357 ).}

\item{ V.~V.~Klimov, Sov. Phys. JETP  {\bf 55} (1982) 199
( Zh.Eksp.Teor.Fiz. {\bf 82} (1982) 336 ).}

\item{ H.~A.~Weldon, Phys. Rev. {\bf D 26} (1982) 2789.}

\item{ E.~J.~Levinson and  D.~H.~Boal, Phys. Rev. {\bf D 31} (1985)
3280.}

\item{ O.~K.~Kalashnikov, Z. Phys. {\bf C 39} (1988) 427.}

\item{ E.~Petitgirard,  Z. Phys.  {\bf C 54} (1992) 673.}

\item{ O.~K.~Kalashnikov, JETP Lett.  {\bf 41} (1985) 582
( Pis'ma Zh.Eksp.Teor.\protect\\ Fiz. {\bf 41} (1985) 477 ).}

\item{ E.~S.~Fradkin, Proc. (Trudy ) P.N.Lebedev Physics Inst. {\bf 29}
(1967) 1.}

\item{ K.~Kajantie and P.~V.~Ruuskaven, Phys.Lett. {\bf B 121 }
(1983) 352.}

\item{ A.~Perez Martinez ,H.~Perez Rojas and A.~Zepeda, Phys.Lett.
{\bf B 366 } (1996) 235; A.~Perez Martinez, A.~Zereda and H.~Perez
Rojas, Int. Jour. Mod. Phys. {\bf 11} (1996) 5093. }

\item{ E.~Braaten and R.~D.~Pisarski, Phys. Rev. Lett. {\bf 64}
(1990) 1338; Nucl. Phys. {\bf B337} (1990) 569.}

\item{ R.~Kobes, G.~Kunstater and A.~Rebhan, Phys. Rev. Lett. {\bf
64} (1990) 292.}

\item{ I.~V.~Chub and V.~V.~Skalozub, to be published in Z. Phys.}

\item{ A.~J.~Gentles and D.~A.~Ross,  SHEP-96-32, hep-ph/9611390.}

\item{ H.~Suganuma, S.~Sasaki and H.~Toki, Nucl. Phys. {\bf B 435}
(1995) 207;  S.~Sasaki, H.~Suganuma  and H.~Toki, Prog. Theor. Phys.
{\bf 94} (1995) 373.}

\end{enumerate}
\end{document}